\documentclass[journal,final,twocolumn,10pt,twoside]{IEEEtran}

\IEEEoverridecommandlockouts

\usepackage[pdftex]{graphicx}
\usepackage{url}
\usepackage{siunitx} 
\usepackage{amsmath,amsfonts,amssymb,amsthm}
\usepackage{gensymb}

\usepackage[noadjust]{cite}
\usepackage[caption=false,font=footnotesize]{subfig}  
\usepackage{relsize}
\usepackage{makecell}
\usepackage{array}
\usepackage{lipsum}

\usepackage{multicol}

\usepackage{todonotes} 
\definecolor{morange}{rgb}{0.8,0.2,0}
\definecolor{mblue}{rgb}{0,0.3,1.0}
\definecolor{mred}{rgb}{0.9,0.1,0.1}
\definecolor{mpurple}{rgb}{0.5 0.1 0.7}

\begin{document}

\title{\huge Simulation Study and Analysis of Diffusive Molecular Communications with an Apertured Plane}
\author{Mustafa~Can~Gursoy,~\IEEEmembership{Student~Member,~IEEE,}
	H.~Birkan~Yilmaz,~\IEEEmembership{Member,~IEEE,}
	Ali~E.~Pusane,~\IEEEmembership{Senior~Member,~IEEE,}
	and~Tuna~Tugcu,~\IEEEmembership{Member,~IEEE} 
	\thanks{M. C. Gursoy was with the Department of Electrical and Electronics Engineering, Bogazici University. He is now with the Ming Hsieh Department of Electrical and Computer Engineering, University of Southern California, 90089, Los Angeles, CA, USA (e-mail: mgursoy@usc.edu).}
	\thanks{H. Birkan Yilmaz and T. Tugcu are with the Department of Computer Engineering, NETLAB, Bogazici University, 34342, Istanbul, Turkey (e-mails:  \{birkan.yilmaz, tugcu\}@boun.edu.tr).}
	\thanks{A. E. Pusane is with the Department of Electrical and Electronics Engineering, Bogazici University, 34342, Istanbul, Turkey (e-mail: ali.pusane@boun.edu.tr).}
}

\markboth{Accepted to IEEE Transactions on NanoBioscience}
{Accepted to IEEE Transactions on NanoBioscience}

\maketitle

\begin{abstract}
Molecular communication via diffusion (MCvD) is a method of achieving nano- and micro-scale connectivity by utilizing the free diffusion mechanism of information molecules. The randomness in diffusive propagation is the main cause of inter-symbol interference (ISI) and the limiting factor of high data rate MCvD applications. In this paper, an apertured plane is considered between the transmitter and the receiver of an MCvD link. Either after being artificially placed or occurring naturally, surfaces or volumes that resemble an apertured plane only allow a fraction of the molecules to pass. Contrary to intuition, it is observed that such topology may improve communication performance, given the molecules that can pass through the aperture are the ones that take more directed paths towards the receiver. Furthermore, through both computer simulations and a theoretical signal evaluation metric named signal-to-interference and noise amplitude ratio (SINAR), it is found that the size of the aperture imposes a trade-off between the received signal power and the ISI combating capability of an MCvD system, hinting to an optimal aperture size that minimizes the bit error rate (BER). It is observed that the trend of BER is accurately mirrored by SINAR, suggesting the proposed metric's applicability to optimization tasks in MCvD systems, including finding the optimal aperture size of an apertured plane. In addition, computer simulations and SINAR show that said optimal aperture size is affected by the location of the aperture and the bit rate. Lastly, the paper analyzes the effects of radial and angular offsets in the placement of the apertured plane, and finds that a reduction in BER is still in effect up to certain offset values. Overall, our results imply that apertured plane-like surfaces may actually help communication efficiency, even though they reduce the received signal power.

\end{abstract}


\IEEEpeerreviewmaketitle

\section{Introduction}

\par Molecular communication via diffusion (MCvD) is communication paradigm that enables information transfer by utilizing the diffusive nature of messenger molecules in fluid media-um \cite{birkansurvey}. In an MCvD system, the information can be typically encoded in the quantity \cite{CSKMOSK}, type \cite{isomermosk}, time of release \cite{impulseDirac}, spatial index of emission \cite{bizimMSSKpaper_TCOM}, and other combinations of these physical properties of the molecular wave. 

\par After their emission to the fluid communication environment, the messenger molecules exhibit Brownian motion \cite{kitap}. Brownian motion introduces randomness in the arrivals of the messenger molecules, which causes the MCvD channel characterization to be done through time dependent functions of received molecules \cite{3Dchar}. Furthermore, said randomness in molecule arrivals is also the main cause of the well-known inter-symbol interference (ISI) problem \cite{ISI_new_ref}. ISI is a major contributor to the high error probabilities and low data rate in MCvD systems \cite{MCSK}.

\par Other than inducing ISI, the probabilistic nature of molecule arrivals also causes some emitted molecules to fail in arriving at the receiver. From a communications engineering stand point, this phenomenon corresponds to a channel-caused attenuation of the molecular signal and leads to a low received signal power at the receiver end \cite{antenna_conf}. In addition, in both macro- and micro-scale applications, there may exist artificially placed or naturally present obstacles, which may lead to further loss in received signal strength \cite{weisi_engel}. Said obstacles may also be multiple layered, as in a porous media \cite{weisi_porous}, and cause further attenuation and arrival delay. In order to combat channel attenuation, several works propose increasing the receivers' molecule capture capability by equipping them with reflecting shells \cite{antenna_conf,directional_rec}. Additionally, in biological molecular communication applications, channel attenuation can also occur due to unwanted molecules that bind to the receiver's receptors and block them. In such a case, \cite{alarcon_AD} considers unbinding the blocking molecules using chemical and optogenetic methods.

\par For an MCvD system consisting of a single transmitter and a single absorbing receiver, a recent work has proposed to partially count the arriving molecules at the receiver end \cite{bayramTCOM}. In other words, \cite{bayramTCOM} considers absorbing all molecules that arrive at the receiver as usual, but only counting the ones that hit to a specific region of the receiver's surface. Interestingly, the study shows that if the counted molecules are the ones that arrive at the receiver's side that faces the transmitter, a sense of directivity can be achieved. This directivity reduces the bit error rate (BER), even though the fractional counting effectively reduces received signal power. Another approach that targets a similar physical phenomenon can be found in \cite{tunel_bizimki}, which considers a tunnel-like absorbing boundary that extends from the transmitter towards the receiver and removes/degrades the molecules that hit it. Both \cite{bayramTCOM} and \cite{tunel_bizimki} hint to the ability of achieving desirable performance gains by introducing directivity to the system.

\par Defining \textit{directivity} in an MCvD system as allowing molecules that take shorter paths (and time) to arrive at the receiver and rejecting the others, this paper considers enhancing the communication efficiency of an MCvD system with an apertured plane. The considered apertured plane consists of a fully reflective and practically infinite (sufficiently large) 2-D slab with a circular aperture in the middle. The molecules are able to freely pass if they propagate through the aperture, but get elastically reflected if they hit the plane's body. Considering the aforementioned topology and our earlier preliminary work in \cite{molcom_paper_2019}, the main contributions of this paper can be listed as follows:
\begin{itemize}
    \item Overall, we find that the existence of such an apertured plane provides a similar directivity as defined before.
    \item We confirm that the pseudo-selective passability through the apertured plane indeed decreases the received signal strength. However, since the blocked molecules mostly consist of the ones that take longer paths to arrive at the receiver, we find that an apertured plane provides a very beneficial ISI mitigation. Overall, even though the received signal is attenuated, we point out that the introduced ISI mitigation reduces BER when compared to the scenario without the plane. This finding also exemplifies that even though naturally occurring physical obstacles decrease received signal strength, not all of them are detrimental to the overall communication performance. Furthermore, it hints to the existence of other apertured plane-like physical obstacles that can provide similar directivity, suggesting a further research direction.
    \item Furthermore, we exploit a trade-off between ISI reduction and received signal power (\cite{multrec_absorb}) that is dependent on the aperture size of the apertured plane, which suggests an optimal aperture size in terms of minimizing BER.
    \item We propose a novel theoretical performance metric called signal-to-interference and noise amplitude ratio (SINAR) that can be evaluated easily and accurately mirrors the behavior of BER, which suggests that the optimal aperture size can also be found by maximizing SINAR. Furthermore, we have shown that SINAR serves as a generalization of the well-known signal-to-interference ratio metric (SIR, \cite{burcuSIR}) for non-asymptotic transmission power. Overall, we believe SINAR is a promising cost function that can be utilized for further equalization and other optimization tasks.
    \item We show that the optimal aperture size is dependent on topological and communication parameters, such as the data rate, transmitted molecular signal strength (i.e. emitted number of molecules), and the position of the apertured plane. 
    \item We discuss and acknowledge that a concentric manual/natural placement of such an aperture plane is a difficult task and investigate the performance loss when the perfect alignment is lost. We find out that compared to the scenario without the plane, the existence of the apertured plane still provides BER reduction up to a certain misalignment due to a radial offset.
\end{itemize}

\section{System Model}
\label{sec:systemmodel}

\par In this paper, an MCvD system between a single point transmitter and a spherical receiver with a radius of $r_r$ is considered in an unbounded, driftless, three dimensional communication environment. The receiver in the MCvD link is assumed to be fully absorbing, which causes it to count all molecules that arrive and remove them from the communication environment (i.e., each hitting molecule contributes to the received signal only once). The closest distance between the point transmitter and receiver is denoted as $d$, which allows the point-to-center distance between the transmitter and the receiver ($r_0$) to be $d+r_r$.

\par
\begin{figure}[!t]
	\centering
	\includegraphics[width=0.48\textwidth]{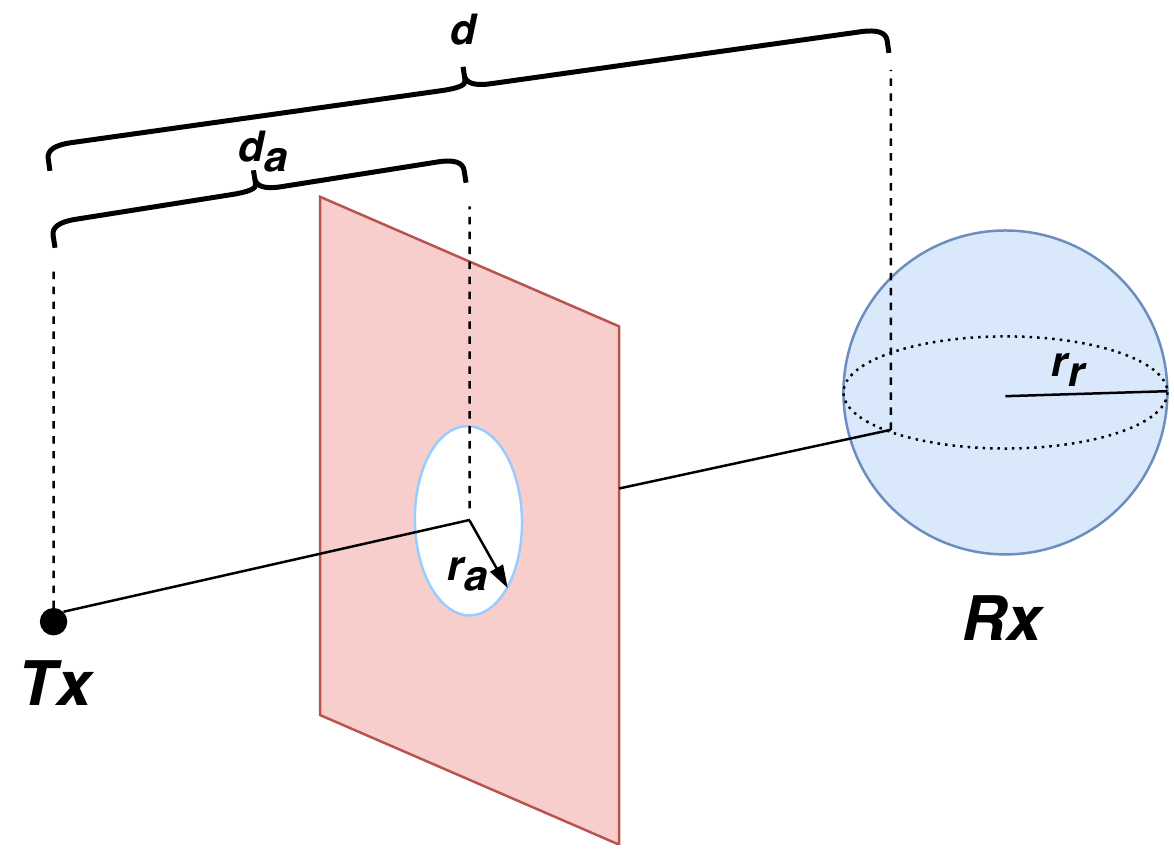}
    \caption{The system model of interest.}	
	\label{fig:systemmodel}
\end{figure}

 Other than the transmitter and receiver nanomachines, the considered system model includes a fully reflective planar surface whose distance to the transmitter is denoted by $d_a$. The surface has a circular aperture with a radius of $r_a$, whose center is aligned with the spherical receiver's center and the point transmitter. Apart from the apertured region, the surface of the apertured plane is assumed to  reflect the molecules that hit its body in an elastic manner. Overall, the considered system model, including the transmitter, the receiver, and the aperture plane, is presented in Fig.~\ref{fig:systemmodel}.

\par In the point-to-absorbing sphere MCvD system without the apertured plane, \cite{3Dchar} introduces the time distribution of a molecule's arrival at the receiver, $f_{hit}(t)$, as 
\begin{equation}\label{eq:arrival_pdf}
f_{hit}(t) = \frac{r_{r}}{r_{0}} \frac{1}{\sqrt[]{4\pi Dt}} \frac{r_{0}-r_{r}}{t} e^{- \frac{(r_{0}-r_{r})^2}{4Dt} }.
\end{equation}
Here, $D$ denotes the diffusion coefficient of the utilized carrier molecule. The time integral of \eqref{eq:arrival_pdf} yields $F_{hit}(t)$, the probability of a single molecule's arrival at the receiver between time $0$ and $t$, which is given by \cite{3Dchar} as:
\begin{equation}\label{eq:arrival_cdf}
F_{hit}(t) = \frac{r_{r}}{r_{0}} \text{erfc}\bigg( \frac{r_{0}-r_{r}}{\sqrt[]{4Dt}} \bigg),
\end{equation}
where $\text{erfc}(\cdot)$ denotes the complementary error function. 

\par The system model of interest, which is presented in Fig. \ref{fig:systemmodel}, includes a reflecting obstructive surface with an aperture, which we refer to as the apertured plane in the manuscript. However, \eqref{eq:arrival_cdf} is only applicable for the system that consists of only a point transmitter and a spherical receiver, and not anything else. Therefore, \eqref{eq:arrival_cdf} cannot be directly used to obtain the $F_{hit}(t)$ function corresponding to the system described in Fig. \ref{fig:systemmodel}. Instead, this paper utilizes a random walk-based Monte Carlo simulation implemented on a custom simulator using MATLAB to obtain $F_{hit}(t)$. To simulate the random movement of the molecules, the position at time $t+\Delta t$ is incremented from the position at time $t$ by performing
\begin{equation}
\begin{split}
    X(t+\Delta t) =& X(t) + \Delta X,\\
    Y(t+\Delta t) =& Y(t) + \Delta Y,\\
    Z(t+\Delta t) =& Z(t) + \Delta Z,
\end{split}
\end{equation}
where all $\Delta X$, $\Delta Y$, and $\Delta Z$ are Gaussian random variables that are defined as $\mathcal{N}(0,2 D \Delta t)$ \cite{kitap}. Here, $D$ denotes the diffusion coefficient of the utilized information molecule in the specific environment.

\par In a time-slotted point-to-absorbing sphere MCvD system, the values of $F_{hit}(t)$ at integer multiples of the symbol duration are equivalent to the probabilities of a single molecule's arrival until the according points in time after its emission. Subtracting consecutive values of these $F_{hit}(t)$ samples yields $h_1,h_2,\dots,h_L$, which are referred to as the channel coefficients of an MCvD system. Here, $L$ denotes the ISI window (channel memory) of the MCvD system at hand. Even though the heavy right tail of \eqref{eq:arrival_pdf} causes the actual channel memory to be infinite, a finite but large $L$ is chosen for practical purposes \cite{BERconverge}. Mathematically, the $k^{th}$ channel coefficient of an MCvD system with symbol duration $t_s$ can be expressed as
\begin{equation}\label{eq:FIRcoefs}
h_k = F_{hit}(kt_s) - F_{hit}((k-1)t_s), \hspace{0.2cm} k = 1,2,\dots,L.
\end{equation}

\par The $k^{th}$ coefficient in \eqref{eq:FIRcoefs} represents the probability of a \textit{single} molecule's arrival in the $k^{th}$ symbol duration after its release. When emitting multiple molecules per release, the number of molecules that arrive at the receiver in the $k^{th}$ receiver is represented with a binomially distributed random variable with success probability $h_k$. Denoting $N_i^{Tx}$ as the number of transmitted molecules and $R_i$ as the number of received molecules at the $i^{th}$ symbol interval, 
\begin{equation}\label{eq:binomial}
R_i = \sum _{k=1}^{L} R_{i,k} 
\end{equation} 
where $R_{i,k}$ denotes the number of received molecules during the $i^{th}$ symbol interval that were emitted at the start of the $k^{th}$ symbol interval, and $R_{i,k} \sim \mathcal{B}(N_{i-k+1}^{Tx},h_{k})$. Following the findings of \cite{arrivalmodel}, \eqref{eq:binomial} can be approximated as a normal random variable given by
\begin{align}\label{eq:receivedGaus}
R_{i} \sim \mathcal{N}  \big(\sum _{k=1}^{L} N_{i-k+1}^{Tx}h_{k},  \sum _{k=1}^{L}N_{i-k+1}^{Tx}h_{k}(1-h_{k})\big).
\end{align}

\par As it is the most widespread modulation scheme in the molecular communication literature \cite{CSKMOSK}, on-off keying-based binary concentration shift keying (BCSK) is utilized in this paper. When transmitting a BCSK symbol, a bit-1 is mapped to emitting $M$ molecules ($N_{i}^{Tx}=M$) from the transmitter and a bit-0 is represented without an emission ($N_{i}^{Tx}=0$) for the symbol duration. The probability of transmitting a bit-0 or a bit-1 is chosen equal ($P_0 = P_1 = 0.5$).

\par At the other end, the receiver considered in this paper collects and counts the absorbed molecules until the end of the symbol duration, and demodulates using a single threshold as
\begin{equation}
    \label{BCSK_demod}
    R_i  \underset{H_0}{\overset{H_1}{\gtrless}} \tau,
\end{equation}
where $\tau$ is the threshold value and $H_\alpha$ is the hypothesis that corresponds to the demodulated bit $\hat{x}_i$ being equal to $\alpha$. Note that the considered time slotted channel assumes synchronization between the transmitter and the receiver, which may be accomplished by a method similar to as presented in \cite{synchronization}.

\section{Effects of the Apertured Plane on System Performance}
\label{sec:BER_SINAR}

\par The apertured plane blocks some molecules' paths by reflecting them, hinders them from arriving at the receiver, and allows some to pass and continue their propagation towards the receiver. Intuitively, the existence of such an obstacle decreases the overall number of arriving molecules at the receiver end \cite{weisi_porous}, which corresponds to a decrease in received signal power. On the other hand, the majority of the molecules that are blocked by the apertured plane are the molecules that take longer paths to arrive at the receiver end, which implies a reduction in ISI. Acknowledging these phenomena as prominent effects on communication efficiency, the overall error performance of the system is hypothesized to be affected by both. Therefore, in order to comment on the overall effect of an apertured plane of communication performance, this section presents BER performances of BCSK modulated bit streams under the topology shown in Fig. \ref{fig:systemmodel}.

\par In the performed computer simulations, the exact topology in Fig. \ref{fig:systemmodel} is generated, and a random walk-based MATLAB simulator as mentioned in Section \ref{sec:systemmodel} is employed to generate $F_{hit}(t)$ and the channel coefficients. Note that for each topology with a different $d_a$, $r_a$, $r_r$, and/or $d$ value, the corresponding $F_{hit}(t)$ is different. Therefore, a different particle-based simulation is performed to generate the appropriate $F_{hit}(t)$ for each different topological parameter set in this paper. Each Monte Carlo simulation is performed for a total duration of $\SI{10}\second$, using $10^5$ molecules and with an incremental time step of $\Delta t = \SI{1e-4}\second$. As the control group, the well-known scenario including a single point transmitter and a single absorbing receiver in an unbounded diffusive environment is considered. Note that for such a topology, $F_{hit}(t)$ can easily be obtained using \eqref{eq:arrival_cdf}. After generating the channel coefficients for both topologies, BER simulations using bit streams of length $10^7$ bits are conducted. Both topologies are evaluated over the well-known BCSK scheme with the decoder as presented in \eqref{BCSK_demod}. 

\par Due to the signal-dependent nature of ISI and noise in MCvD systems \cite{sig-dependent-noise}, the theoretical evaluation of BER necessitates evaluations over all possible bit combinations \cite{lolTSP}. For an MCvD channel with memory $L$, theoretical BER evaluation ideally requires $2^L$ cases' error probability evaluations. Stemming from the complexity of such an operation, various performance metrics including SIR \cite{burcuSIR}, signal-to-interference-and-noise ratio (SINR) \cite{SINR_jamali}, and signal-to-interference difference (SID) \cite{bayramTCOM} have been proposed in the literature. Inspired by the existence of a correlation between BER and the ratio between a molecular signal's mean and standard deviation \cite{MaaFjournal}, we propose an alternative performance metric to evaluate the signal quality in this work. We have named the metric signal-to-interference-and-noise amplitude ratio (SINAR) and defined it as 
\begin{equation}\label{eq:SINAR}
    \textrm{SINAR} = \frac{M h_1}{\sum_{k=2}^{L} M h_k + \sum_{k=1}^{L} \sqrt{M h_k (1-h_k)}}.
\end{equation}
As \eqref{eq:SINAR} also suggests, the numerator of SINAR consists of the intended arrival signal, whilst the denominator consists of ISI and the total standard deviation (intended and ISI). Note that SINAR takes the number of transmitted molecules ($M$) into account, which is a key difference when compared to SIR \cite{burcuSIR} and SID \cite{bayramTCOM} metrics. In fact, recalling the SIR definition in \cite{burcuSIR} as $\frac{h_1}{\sum_{k=2}^{L} h_k}$, it can be seen that in the asymptotic case where $M \rightarrow \infty$, SINAR converges to SIR. This equivalency can be straightforwardly shown as
\begin{equation}\label{eq:SINAR_SIR}
\begin{split}
\lim_{M \rightarrow \infty} \textrm{SINAR} =& \lim_{M \rightarrow \infty} \frac{h_1}{\sum_{k=2}^{L} h_k + \frac{1}{\sqrt{M}} \sum_{k=1}^{L} \sqrt{h_k (1-h_k)}} \\
=& \frac{h_1}{\sum_{k=2}^{L} h_k}\\
=& \textrm{SIR}.
\end{split}
\end{equation}
In other words, \eqref{eq:SINAR} and \eqref{eq:SINAR_SIR} imply that SINAR can be thought of as the generalization of SIR into non-asymptotic power cases. Lastly, it is worth mentioning that the fact that $M$ is used in first order makes physical sense, since the number of transmitted molecules is considered to be directly proportional to energy required to synthesize and/or emit them \cite{energymodel}. 

\par
\begin{figure}[!t]
	\centering
	\includegraphics[width=0.48\textwidth]{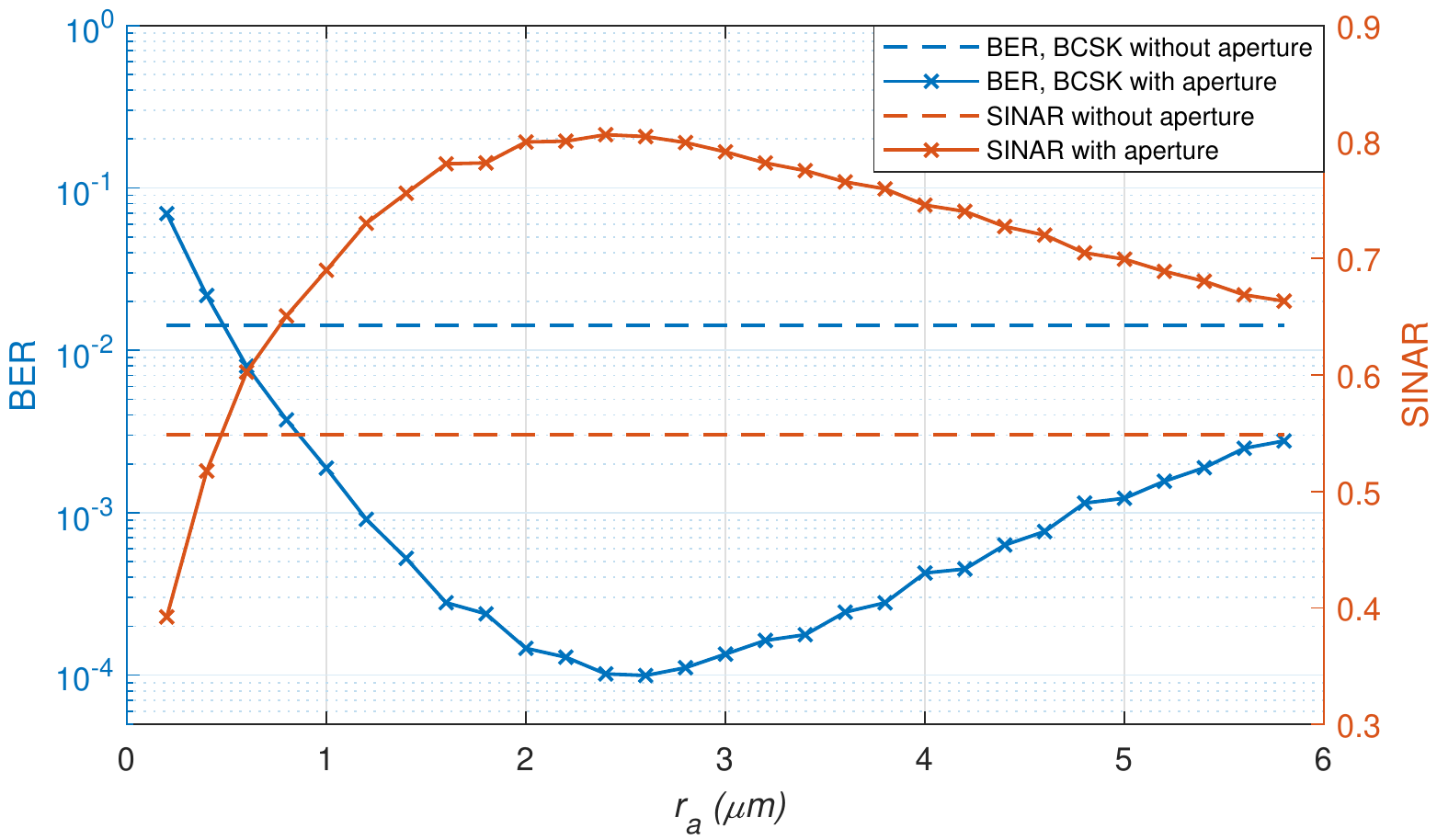} 
	\caption{BER and SINAR vs. $r_a$ curves for BCSK. $M = 1500$ molecules, $t_s = \SI{0.2}\second$, $d = \SI{5}{\micro\meter}$, $r_r = \SI{5}{\micro\meter}$, $d_a = \SI{3}{\micro\meter}$, $D = 79.4 \frac{\SI{}{\micro\meter\squared}}{\SI{}{\second}}$.}
	\label{fig:ra_exp}
\end{figure}

 In order to evaluate the effects of the apertured plane on communication efficiency, BER vs. $r_a$ and SINAR vs. $r_a$ curves are presented in Fig. \ref{fig:ra_exp}. Throughout the paper, including Fig. \ref{fig:ra_exp}, $r_r = \SI{5}{\micro\meter}$, $d = \SI{5}{\micro\meter}$, and $D = 79.4 \frac{\SI{}{\micro\meter\squared}}{\SI{}{\second}}$ are chosen as default values of their corresponding parameters and are taken as such unless specified.

\par Firstly, the results of Fig. \ref{fig:ra_exp} show that SINAR is a good metric to evaluate the performance of an MCvD system, as it reliably mirrors the trend of the BER curve. In other words, Fig. \ref{fig:ra_exp} suggests maximizing SINAR yields close-to-optimal results in terms of minimizing BER, which makes it promising for its use as a cost function in further equalizer and other filter designs. Please note that the computational complexity for evaluating SINAR is much less when compared to evaluating BER.

\par As both the local minimum of BER vs. $r_a$ and the local maximum of SINAR vs. $r_a$ curves in Fig. \ref{fig:ra_exp} suggest, there exists an optimal $r_a$ value that minimizes BER for an MCvD system with given topological and communication parameters. Calling this optimal $r_a$ value as $r_a^*$, Fig. \ref{fig:ra_exp} suggests that for the system with considered parameters, $r_a^*\approx \SI{2.4}{\micro\meter}$. Furthermore, mapping the results of Fig. \ref{fig:ra_exp} to the first paragraph of Section \ref{sec:BER_SINAR}, it can be inferred that the hypothesized trade-off between received signal power and ISI reduction is indeed true. Note that for large $r_a$, the ISI mitigation is minimal and the BER-reducing effect of the apertured plane is less significant. Naturally, as $r_a \rightarrow \infty$, the BER and SINAR curves are expected to approach their \textit{without aperture} counterparts. At the other extreme, an apertured plane with a very small $r_a$ only allows the molecules that take very directed paths towards the receiver, hence providing excellent ISI mitigation. However, this reduction in ISI does not translate to BER reduction since the received signal power becomes too low (too few molecules arrive at the receiver), which makes the received signal weaker against arrival variance (noise). Overall, the optimal value $r_a^*$ provides the best trade-off between the received signal power and ISI mitigation capability, and yields the lowest BER, as also quantified by the SINAR curve.

\par
\begin{figure}[!t]
	\centering
	\includegraphics[width=0.48\textwidth]{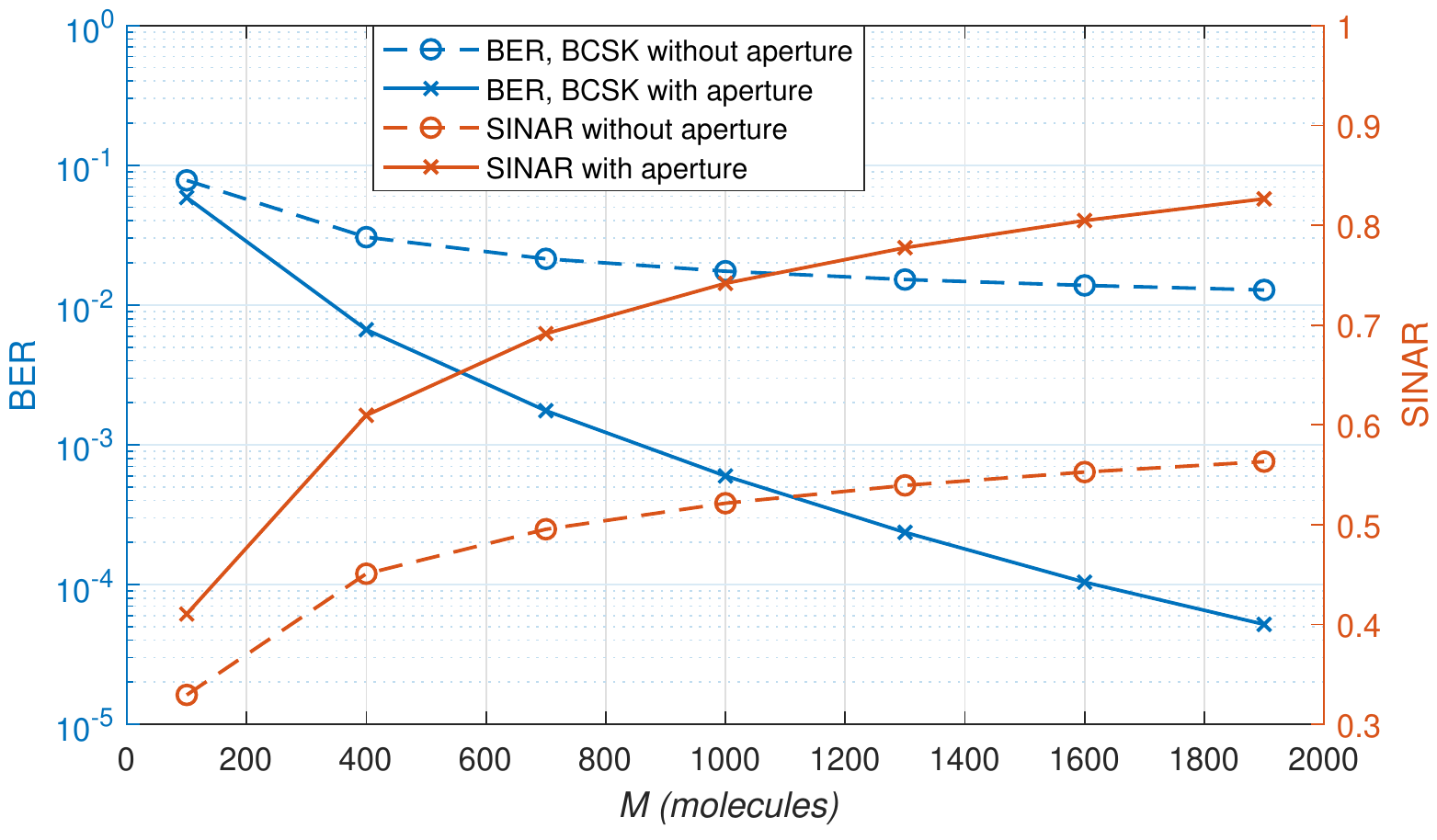} 
	\caption{BER and SINAR vs. $M$ curves for BCSK. $t_s = \SI{0.2}\second$, $d = \SI{5}{\micro\meter}$, $r_r = \SI{5}{\micro\meter}$, $d_a = \SI{3}{\micro\meter}$, $r_a = \SI{2.4}{\micro\meter}$, $D = 79.4 \frac{\SI{}{\micro\meter\squared}}{\SI{}{\second}}$.}	
	\label{fig:M_exp}
\end{figure}

 Another finding of Fig. \ref{fig:ra_exp} is that at $r_a^*\approx \SI{2.4}{\micro\meter}$, a BER gain of roughly two orders of magnitude can be obtained at $t_s = \SI{0.2}\second$ and $M = 1500$ molecules. Referring to the widespread use of BER vs. $M$ curves in molecular communication literature, BER and SINAR vs. $M$ curves at $r_a^* = \SI{2.4}{\micro\meter}$ are presented in Fig. \ref{fig:M_exp}. 

\par The results of Fig. \ref{fig:M_exp} show that the apertured plane not only reduces overall BER, but it also provides a steeper downward slope in the BER vs. $M$ curve. This desirable property can be accounted to the ISI mitigating property of the apertured system.

\section{System Parameters and the Optimal Aperture Size}
\label{sec:ts_heatmap}

\par The results presented in Section \ref{sec:BER_SINAR} show that even though the presence of an apertured plane attenuates received signal power \cite{weisi_engel}, the overall communication efficiency may increase when the aperture allowing the communication is in a certain range in size. Fig. \ref{fig:ra_exp} shows that for a certain parameter set, the trade-off between the received signal power and ISI combating leads to the existence of an optimal aperture radius $r_a^*$. This section investigates the factors that affect $r_a^*$ and $r_a^*$'s behavior with respect to them.

\subsection{Optimal Aperture Radius With Respect to $t_s$}
\par
\begin{figure}[!t]
	\centering
	\includegraphics[width=0.48\textwidth]{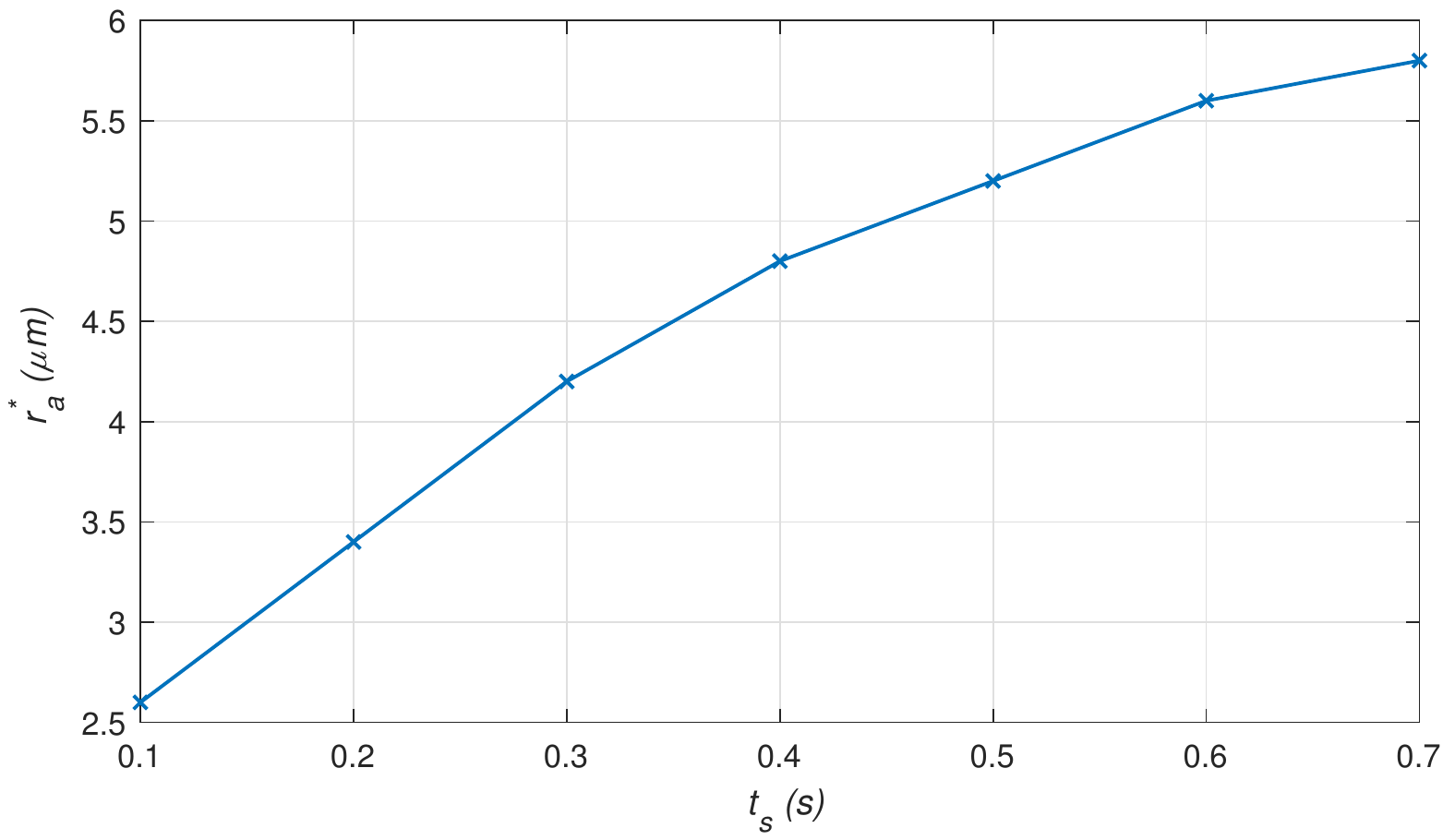} 
	\caption{$r_a^{*}$ vs. $t_s$ curve for BCSK. $M = 100$ molecules, $d = \SI{5}{\micro\meter}$, $r_r = \SI{5}{\micro\meter}$, $d_a = \SI{3}{\micro\meter}$, $D = 79.4 \frac{\SI{}{\micro\meter\squared}}{\SI{}{\second}}$.}	
	\label{fig:tb_exp}
\end{figure}

 Symbol duration is a key parameter of any MCvD system, including the ones involving apertured plane-like obstacles. As noted many times in the molecular communications literature \cite{birkansurvey}, the symbol duration is directly linked to the ISI faced by an MCvD system. Acknowledging the key importance of the received signal power vs. ISI trade-off for the system at hand, Fig. \ref{fig:tb_exp} is presented to illustrate $r_a^*$'s relationship with $t_s$.

\par The results of Fig. \ref{fig:tb_exp} suggest that $r_a^*$ increases with increasing $t_s$. Note that under the same topological parameters, a higher $t_s$ means lower ISI. This phenomenon suggests that the system is better off with a larger $r_a$ allowing more molecules to pass (higher received power), rather than further mitigating the already low ISI.

\subsection{Optimal Aperture Radius With Respect to the Number of Transmitted Molecules}

\par
\begin{figure}[!t]
	\centering
	\includegraphics[width=0.48\textwidth]{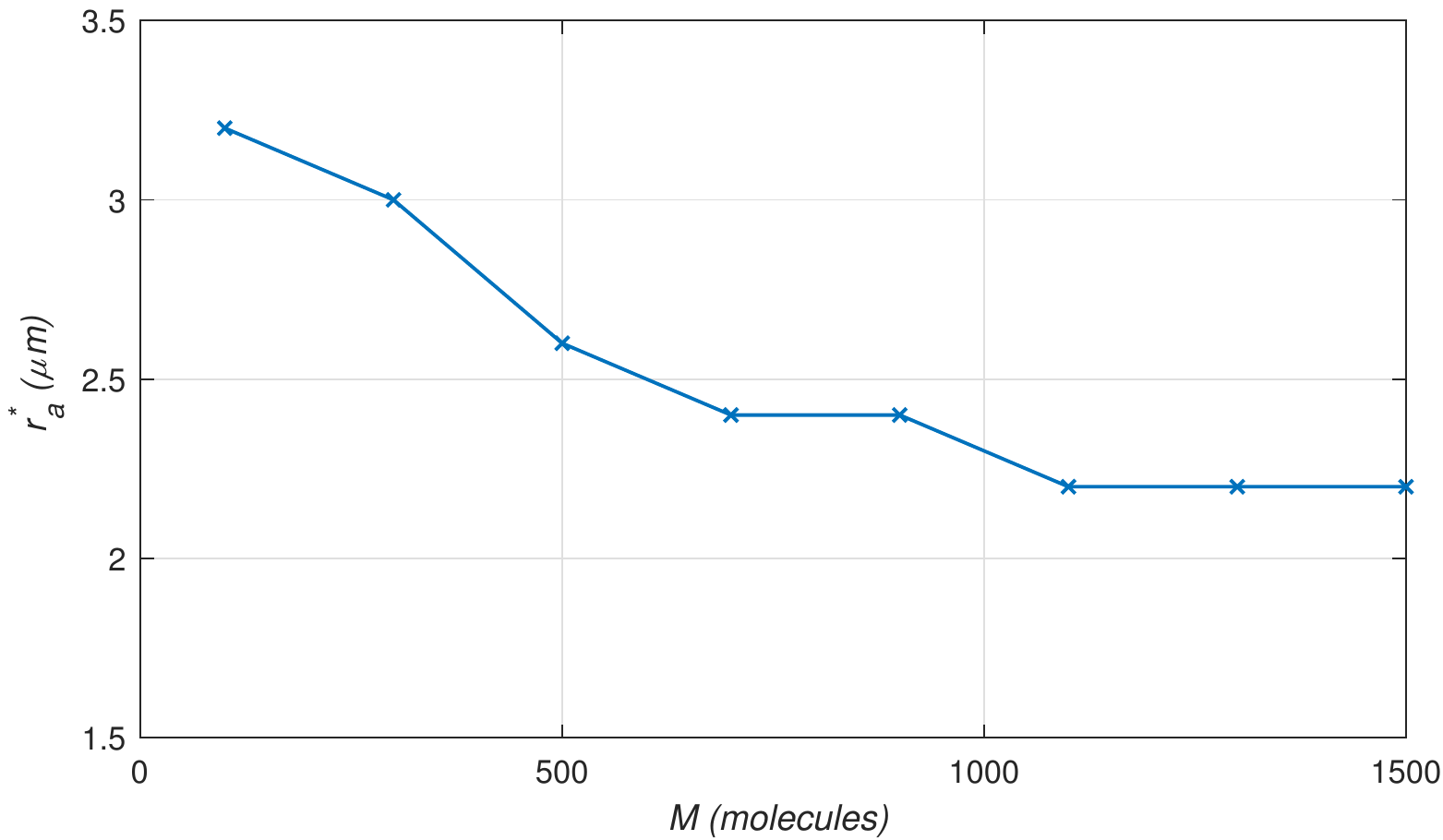} 
	\caption{$r_a^{*}$ vs. $M$ curve for BCSK. $t_s = \SI{0.2}{\second}$, $d = \SI{5}{\micro\meter}$, $r_r = \SI{5}{\micro\meter}$, $d_a = \SI{2.4}{\micro\meter}$, $D = 79.4 \frac{\SI{}{\micro\meter\squared}}{\SI{}{\second}}$.}	
	\label{fig:M_opt}
\end{figure}

 Another key parameter of an MCvD system is the number of transmitted molecules. Recalling that the employed BCSK modulation does not emit any molecules to transmit a bit-0, the average number of transmitted molecules is represented solely by $M$ in this paper. In order to investigate $r_a^*$'s relationship with $M$, Fig. \ref{fig:M_opt} is presented.

\par The results of Fig. \ref{fig:M_opt} show that $r_a^*$ decreases with increasing $M$. Since transmitting more molecules corresponds to a higher transmit power, the system with a larger $M$ is able to handle a more severe signal attenuation and still provide a decent signal power at the receiver end. Noting a small aperture has better ISI combating capability, one can conclude that an MCvD system with higher $M$ is better off with a smaller $r_a$, which explains the downward trend in the $r_a^*$ vs. $M$ curve. We also observe that the downward trend slows down while increasing $M$, and after $M>1000$, the change in $r_a^*$ is negligible (i.e., $r_a^*$ converges to a value when we increase $M$). 

\subsection{Optimal Aperture Radius and the Position of the Apertured Plane}

\par
\begin{figure*}[!t]
	\centering
	\subfloat[$t_s = \SI{0.2}{\second}$, $d = \SI{5}{\micro\meter}$.]{\includegraphics[width=.48\textwidth]{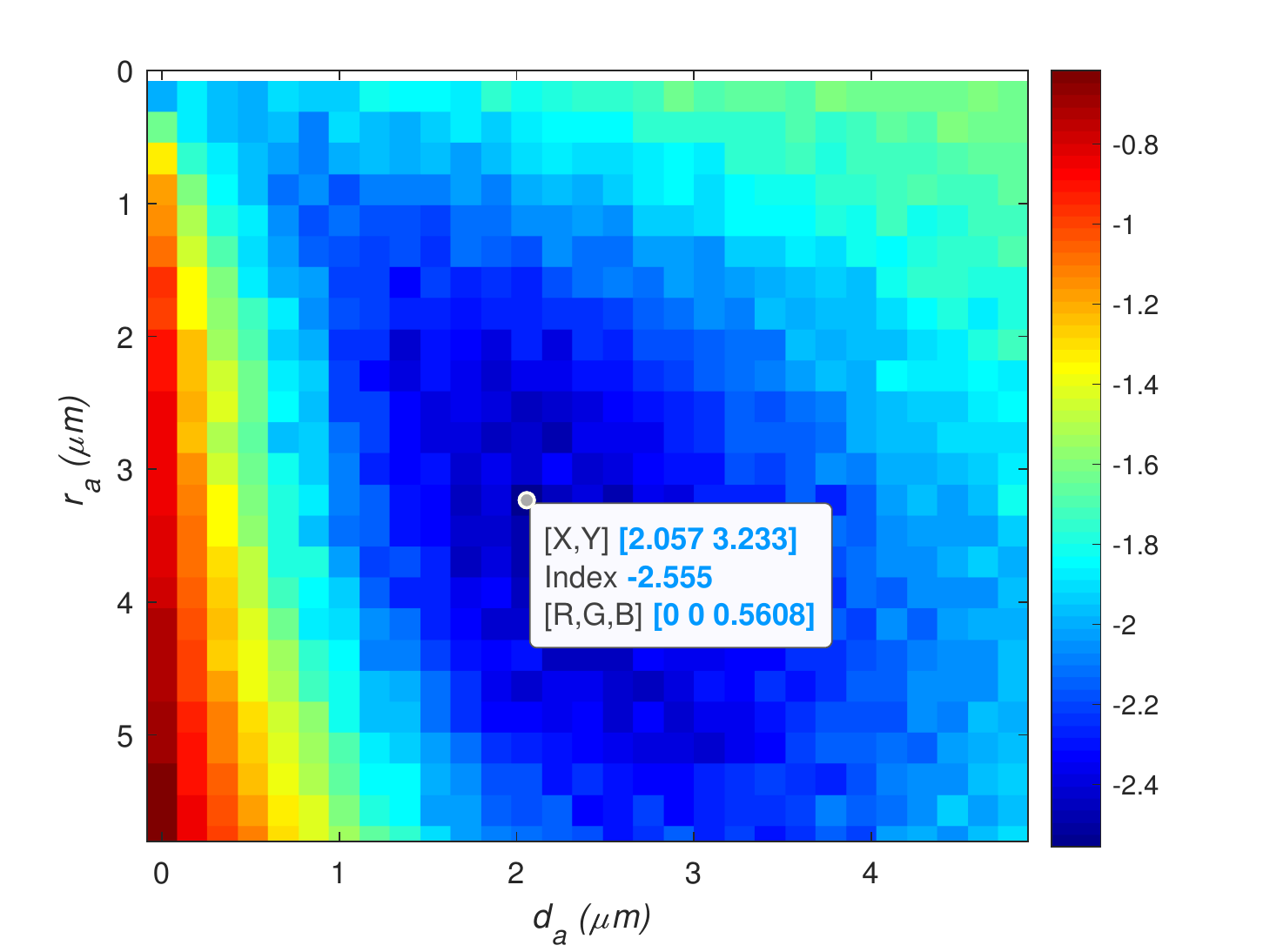}} 
	\subfloat[$t_s = \SI{0.2}{\second}$, $d = \SI{5}{\micro\meter}$.]{\includegraphics[width=.48\textwidth]{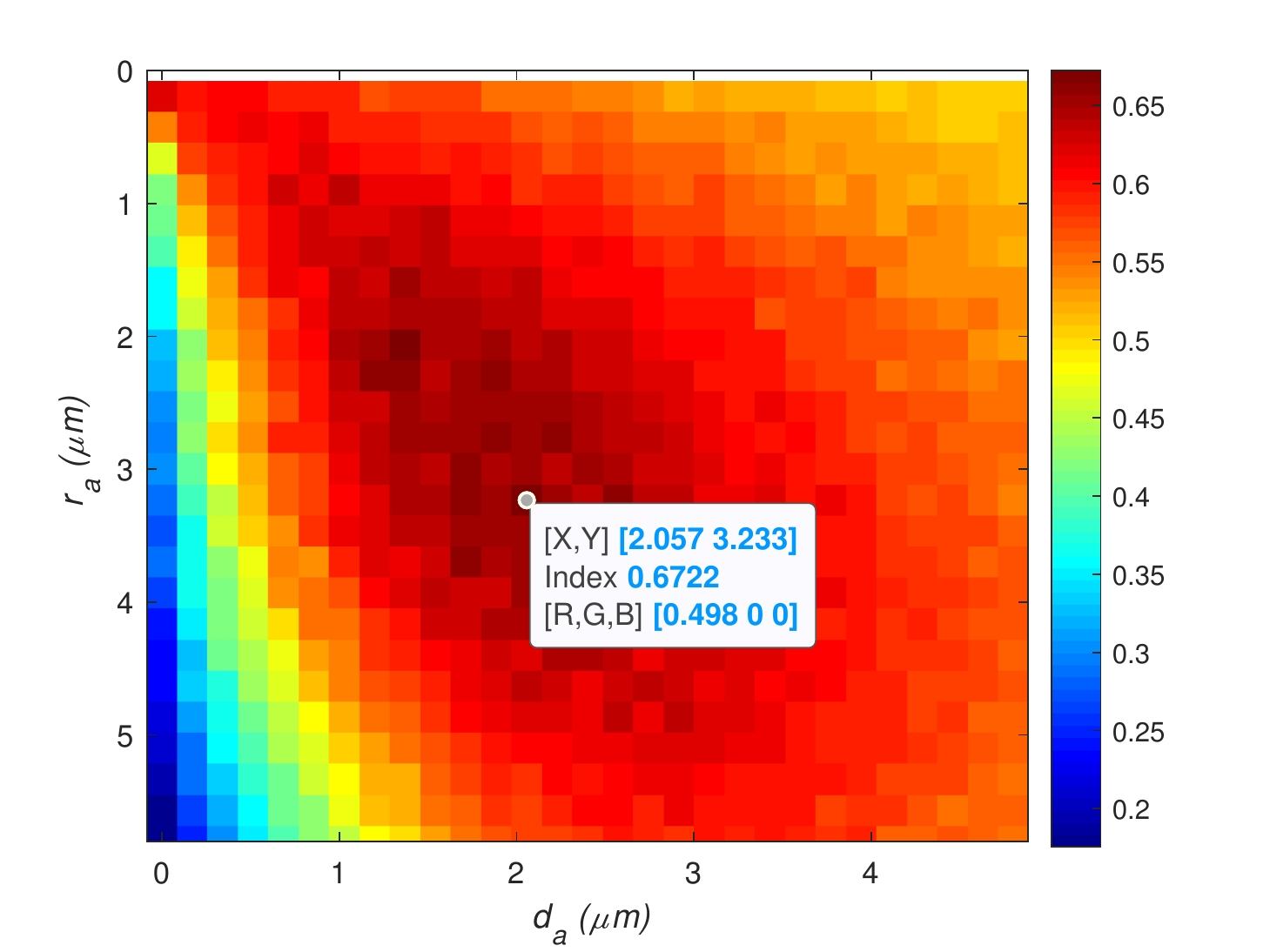}} \quad
	\subfloat[$t_s = \SI{0.4}{\second}$, $d = \SI{7}{\micro\meter}$.]{\includegraphics[width=.48\textwidth]{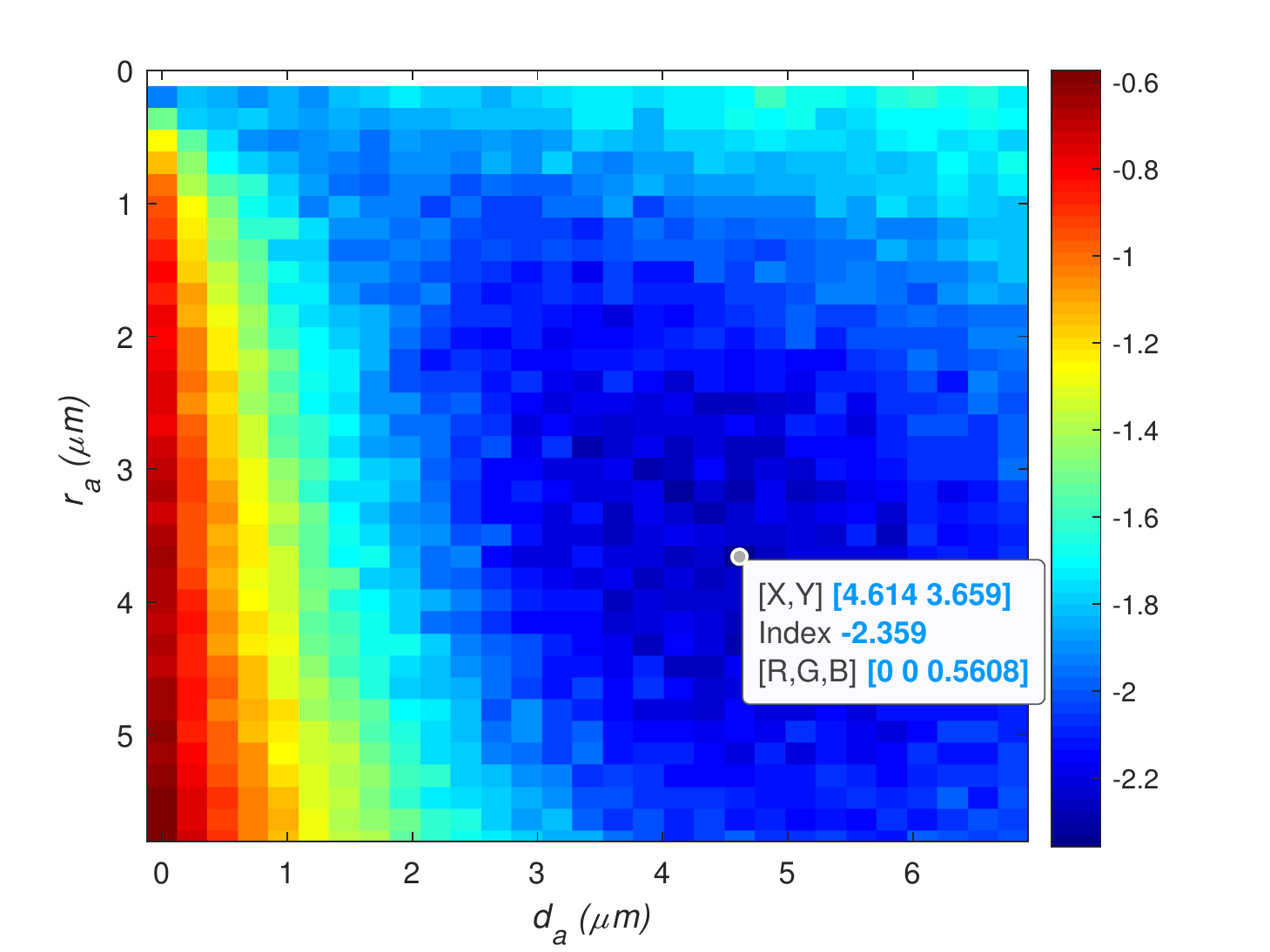}}
	\subfloat[$t_s = \SI{0.4}{\second}$, $d = \SI{7}{\micro\meter}$.]{\includegraphics[width=.48\textwidth]{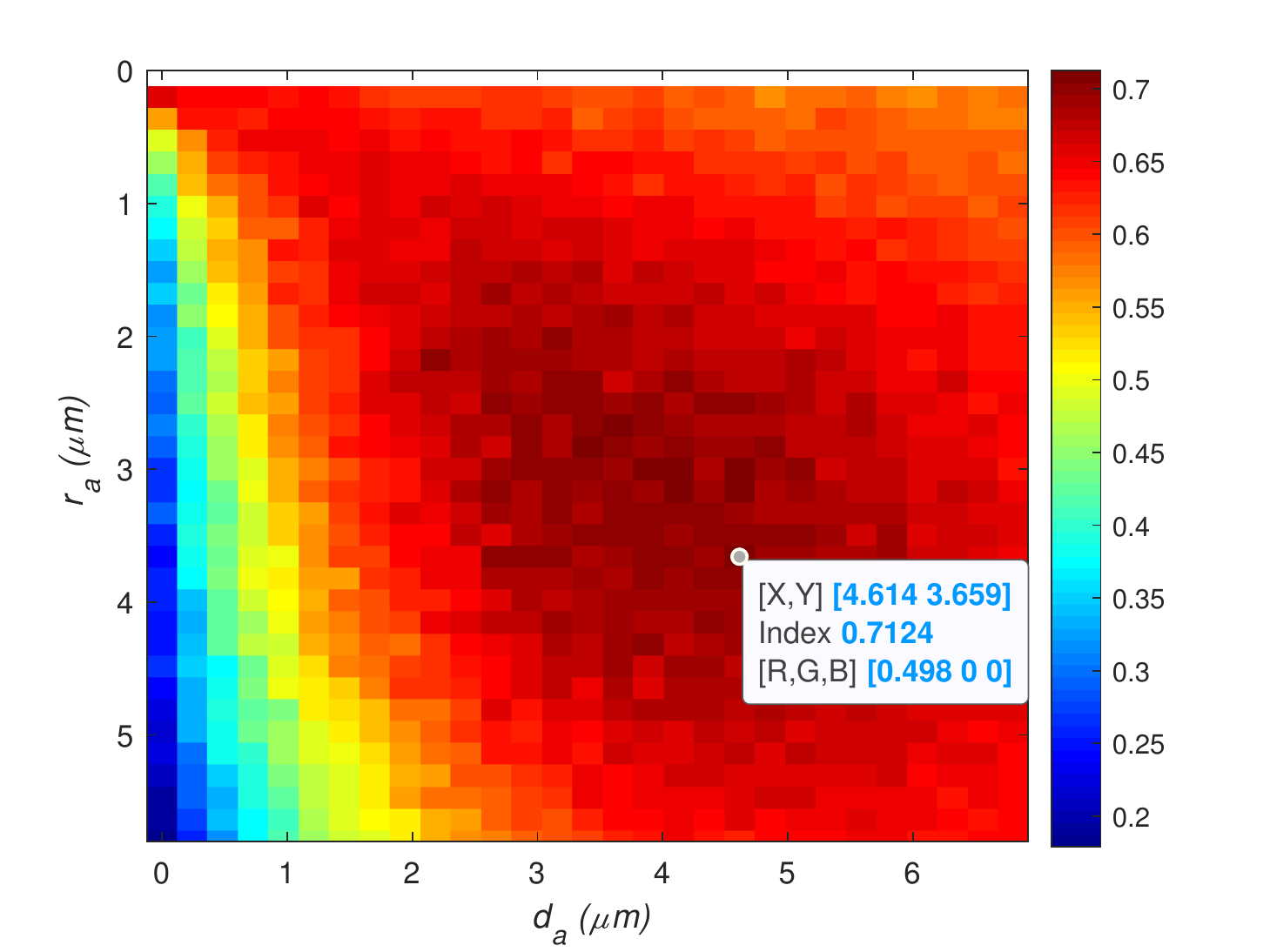}}\\
	\caption{$\log_{10}(\text{BER})$ (left column) and SINAR (right column) vs. $r_a$ and $d_a$ heatmaps for BCSK. $M = 500$ molecules, $r_r = \SI{5}{\micro\meter}$, $D = 79.4 \frac{\SI{}{\micro\meter\squared}}{\SI{}{\second}}$.} 
	\label{fig:heatmapfigure_R1}
\end{figure*}

 This subsection investigates the effects of the position of the apertured plane on $r_a^*$ and investigates the global minimum of BER (or global maximum of SINAR) with respect to $d_a$ and $r_a$. To serve this purpose, Fig. \ref{fig:heatmapfigure_R1}, which holds the $\log_{10}(\text{BER})$ and SINAR vs. $r_a$ and $d_a$ heatmaps for $d = \SI{5}{\micro\meter}$ (Subfigs. a and b) and $d = \SI{7}{\micro\meter}$ (Subfigs. c and d), is presented.

\par To generate Fig. \ref{fig:heatmapfigure_R1}, $F_{hit}(t)$ functions corresponding to each ($d_a,r_a$) pair within their respective sets of $\{0,0.2,0.4,\dots,d-0.2\}\SI{}{\micro\meter}$ and $\{0,0.2,0.4,\dots,5.8\}\SI{}{\micro\meter}$, are obtained. The acquired $F_{hit}(t)$ functions are used in Monte Carlo BER simulations (with $10^6$ trials for each scenario) and analytical SINAR calculations. Note that each $2$-D square bin represents a $\SI{0.2}{\micro\meter} \times \SI{0.2}{\micro\meter}$ grid.

\par From both result sets corresponding to $d = \SI{5}{\micro\meter}$ and $d = \SI{7}{\micro\meter}$, it can be observed that as $d_a$ increases, the $r_a^*$ that corresponds to the corresponding $d_a$ increases as well. This behavior is mainly due to the aforementioned trade-off between the directivity effect (ISI combating) and received signal power that is brought in by the apertured plane. Recall from Fig. \ref{fig:systemmodel} that when $d_a$ is small, the apertured plane is closer to the transmitter. In the small $d_a$ region, it makes more sense to have a small $r_a$ to provide directivity by allowing only more \textit{directed} molecules to pass, since a large $r_a$ value does not provide sufficient blockage, hence ISI combating, near the transmitter. On the other hand, when the apertured plane is closer to the receiver (large $d_a$ region), having a small $r_a$ is equivalent to allowing a very small portion of the molecules to be absorbed, which corresponds to having a very low received signal power. The need to avoid such low received powers causes the system to trade some ISI combating capability by having a larger aperture, increasing $r_a^*$ for larger $d_a$ values. Lastly, it can also be observed from Fig. \ref{fig:heatmapfigure_R1} that the SINAR metric accurately represents the overall BER behavior, and enables the prediction of the optimal $r_a$ and $d_a$ combination analytically.

\section{Performance Under Imperfect Alignment}

\subsection{Performance Under a Radial Offset}

\par Up to this section of the paper, the simulations and analyses consider a perfect alignment between the point transmitter, the center of the circular aperture, and the center of the spherical receiver. However, for scenarios with either manually placed apertured planes or naturally occurring apertured plane-like obstacles, ensuring a perfectly concentric alignment is difficult. This alignment problem is especially an issue for nano and micro-scale MCvD systems. 

\par
\begin{figure}[!t]
	\centering
	\includegraphics[width=0.48\textwidth]{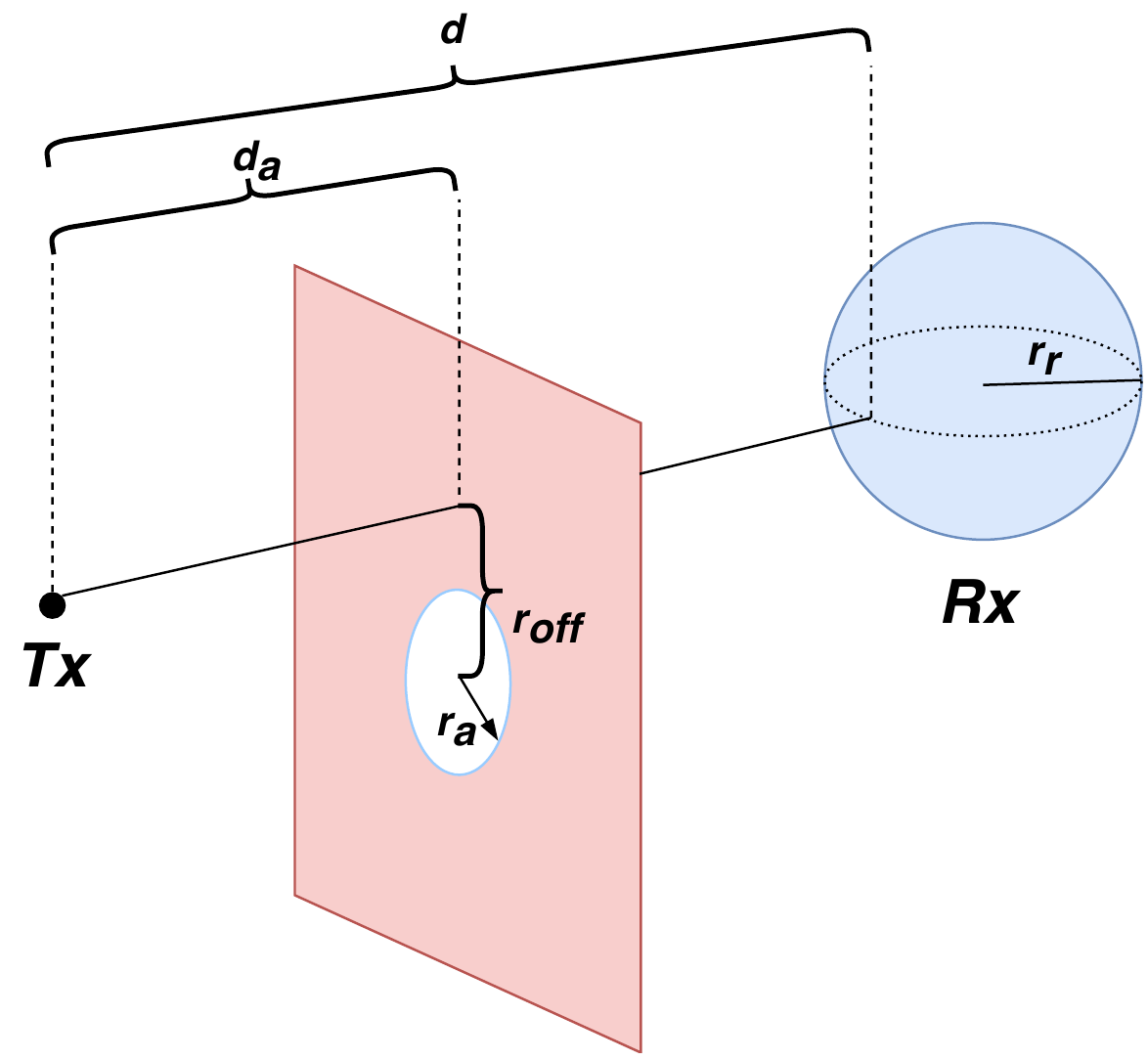} 
    \caption{The apertured plane MCvD system model with offset distance $r_{\textrm{off}}$. $r_{\textrm{off}} = 0$ corresponds to the concentric case.}	
	\label{fig:offsetli_model}
\end{figure}

 Motivated by the alignment problem, this section presents the error performance profile of an MCvD system with an apertured plane that is imperfectly aligned. Fig. \ref{fig:offsetli_model} presents the non-concentric version of Fig. \ref{fig:systemmodel}, where the center of the aperture is $r_{\textrm{off}}$ away from the line that connects the point transmitter and the center of the spherical receiver. Note that $r_{\textrm{off}} = 0$ corresponds to the perfect alignment scenario studied in earlier sections.

\par
\begin{figure}[!t]
	\centering
	\includegraphics[width=0.48\textwidth]{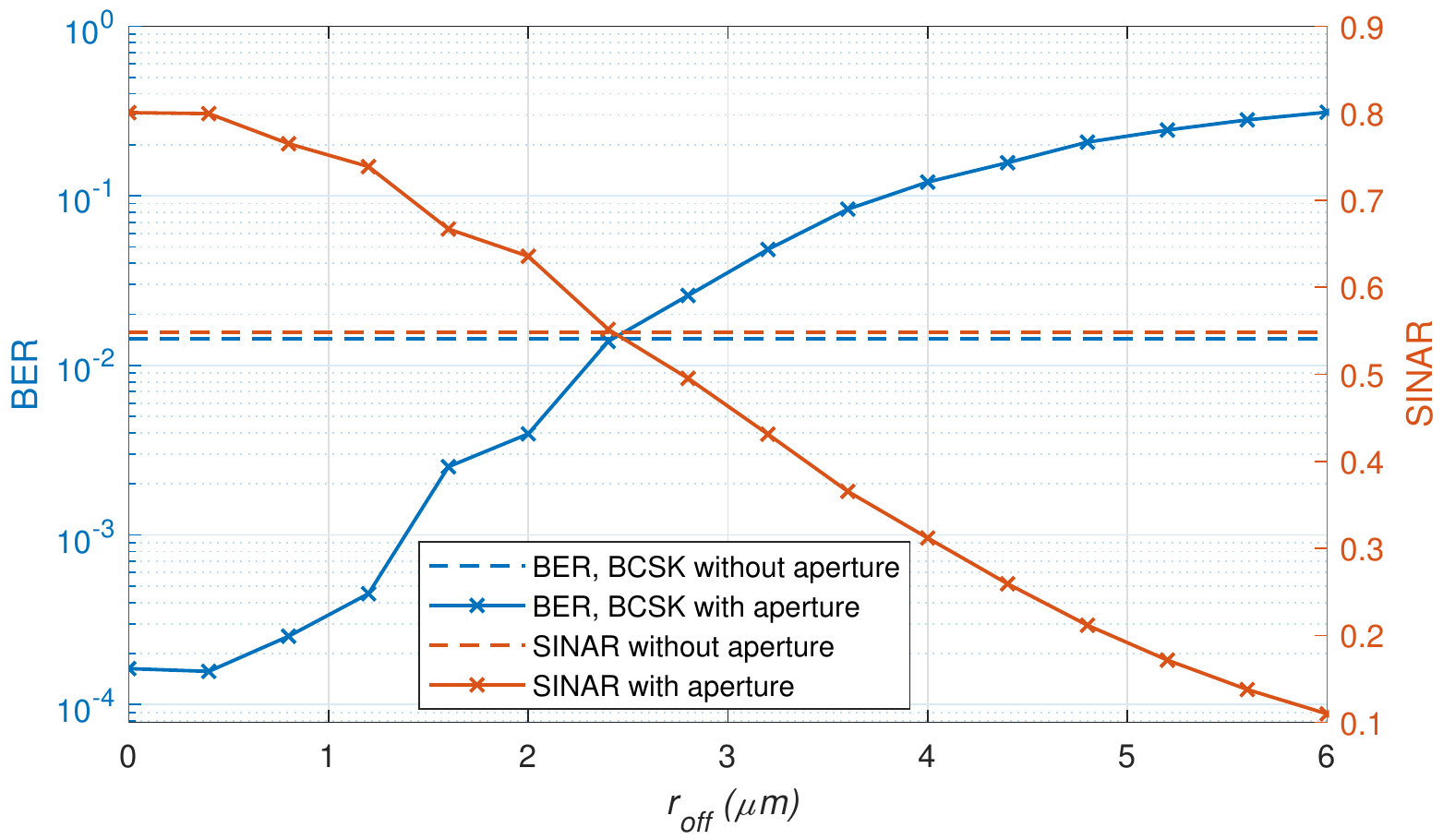} 
	\caption{BER and SINAR vs. $r_{\textrm{off}}$ curves for BCSK. $M = 1500$ molecules, $t_s = \SI{0.2}\second$, $d = \SI{5}{\micro\meter}$, $r_r = \SI{5}{\micro\meter}$, $d_a = \SI{3}{\micro\meter}$, $r_a = \SI{2}{\micro\meter}$, $D = 79.4 \frac{\SI{}{\micro\meter\squared}}{\SI{}{\second}}$. }	
	\label{fig:roff_exp}
\end{figure}

 Fig. \ref{fig:roff_exp} presents the comparative BER and SINAR curves of MCvD systems with and without the apertured plane, where the offset distance $r_{\textrm{off}}$ is used as the swept parameter. Similar to the perfectly concentric scenarios, the $F_{hit}(t)$ functions and the channel coefficients of the imperfectly aligned systems are also generated using Monte Carlo simulations with $10^5$ molecules, for a total duration of $\SI{10}\second$ and an incremental time step of $\Delta t = \SI{1e-4}\second$. After generating the required data, the Monte Carlo BER simulation is performed using bit streams with length $10^6$, in order to ensure a minimum of roughly $100$ bit errors.

\par As the results of Fig. \ref{fig:roff_exp} also confirm intuition, an increasing misalignment $r_{\textrm{off}}$ increases BER for the apertured plane scenario. The reason for this behavior can be accounted to the increased distance the molecules have to cover before arriving at the receiver. The longer distance increases the expected arrival time of the molecules, making the system more prone to ISI. However, it is worth mentioning that the beneficial effect of the apertured plane is still in place up to a certain $r_{\textrm{off}}$ value. For the system with parameters in the caption of Fig. \ref{fig:roff_exp}, the system with a misaligned apertured plane still yields a lower BER than the case without the plane at all up to $r_{\textrm{off}} \approx \SI{2.4}{\micro\meter}$, which is also confirmed by the SINAR metric.

\subsection{The Effect of Aperture Radius Under Imperfect Alignment}

\par
\begin{figure}[!t]
	\centering
	\includegraphics[width=0.48\textwidth]{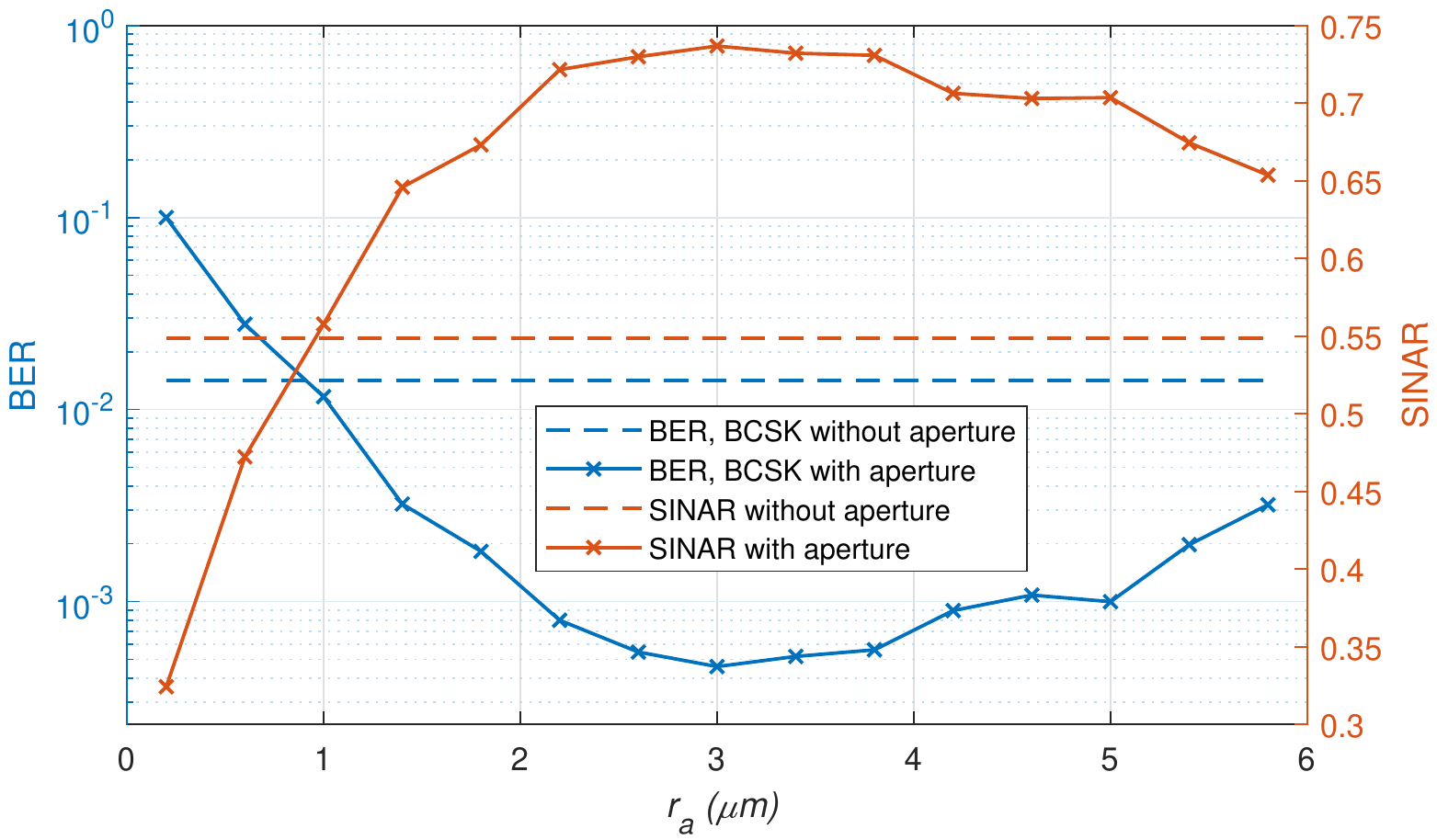} 
	\caption{BER and SINAR vs. $r_{\textrm{a}}$ curves for BCSK. $M = 1500$ molecules, $t_s = \SI{0.2}\second$, $d = \SI{5}{\micro\meter}$, $r_r = \SI{5}{\micro\meter}$, $d_a = \SI{3}{\micro\meter}$, $r_\textrm{off} = \SI{1.5}{\micro\meter}$, $D = 79.4 \frac{\SI{}{\micro\meter\squared}}{\SI{}{\second}}$.}
	\label{fig:ra_exp_sabitroff}
\end{figure}

 Fig. \ref{fig:ra_exp} suggests that in the presence of an apertured plane between the transmitter and the receiver, there exists an optimal aperture size that maximizes SINAR and minimizes BER. Recalling that this non-trivial behavior is found to be stemming from a trade-off between ISI combating and received signal power, this subsection re-addresses the optimal aperture radius problem under imperfect alignment. Fig. \ref{fig:ra_exp_sabitroff} presents the BER vs. $r_a$ curve with a non-zero radial offset ($r_\textrm{off} = \SI{1.5}{\micro\meter}$).

\par The results of Fig. \ref{fig:ra_exp_sabitroff} show that the optimal aperture radius $r_a^*$ shifts to the right as $r_\textrm{off}$ increases. Furthermore, it is found that an increase in $r_\textrm{off}$ also causes an increase in the $r_a$ value that yields an equivalent performance to the no apertured plane scenario. Recall that since an apertured plane only allows molecules to pass through its circular aperture, an imperfect alignment causes the molecules to take longer paths before arriving at the receiver and increases the expected arrival time. On the other hand, for a fixed $r_a$, an offset also causes fewer molecules to be able to arrive at the receiver. In our studies, we have observed that as $r_\textrm{off}$ increases, the system is better-off by partially mitigating the received power loss by having a larger aperture, rather than providing an imperfect directivity. As a result, the BER vs. $r_a$ curve expands rightwards. Overall, it is noteworthy that, although at a lower gain and with a different $r_a^*$ than the concentric case, the communication performance can still be considerably enhanced by an apertured plane in a non-concentric scenario.

\subsection{Performance Under Angular Offsets}

\par
\begin{figure}[!t]
	\centering
	\includegraphics[width=0.48\textwidth]{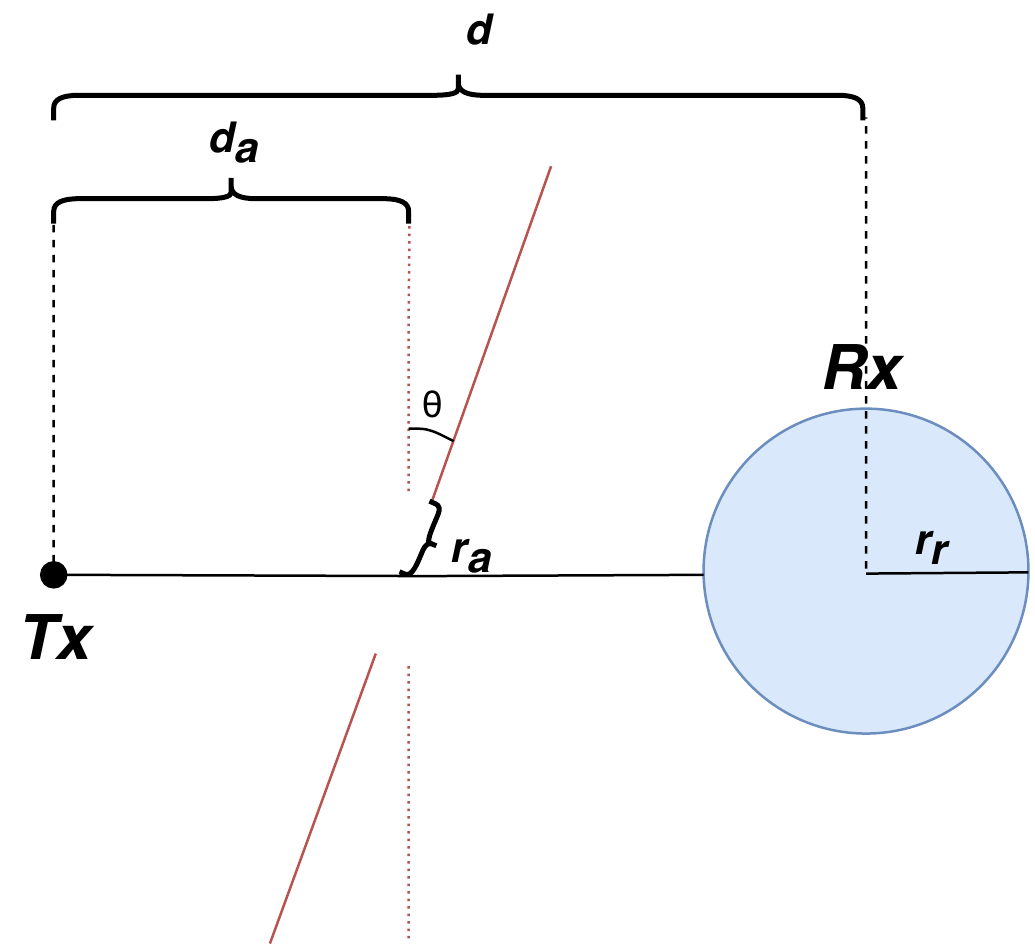} 
    \caption{The (concentric) apertured plane MCvD system model with an angular rotation $\theta$. $\theta = 0$ corresponds to the standard topology.}
	\label{fig:rotation_topology}
\end{figure}

\begin{figure}[!t]
	\centering
	\includegraphics[width=0.48\textwidth]{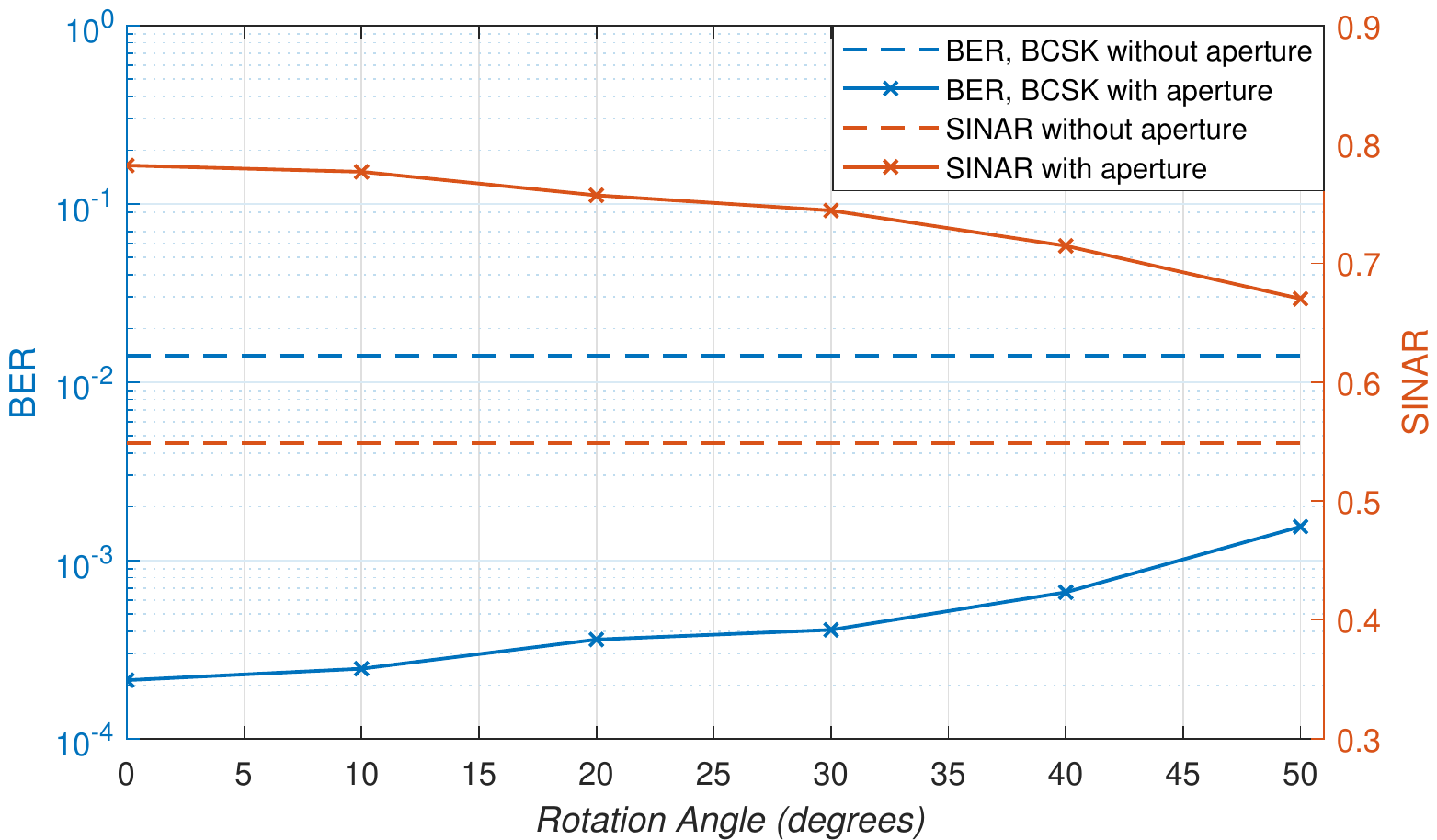} 
	\caption{BER and SINAR vs. $\theta$ curves for BCSK. $M = 1500$ molecules, $t_s = \SI{0.2}\second$, $d = \SI{5}{\micro\meter}$, $r_r = \SI{5}{\micro\meter}$, $d_a = \SI{2}{\micro\meter}$, $r_a = \SI{3}{\micro\meter}$, $r_\textrm{off} = \SI{0}{\micro\meter}$, $D = 79.4 \frac{\SI{}{\micro\meter\squared}}{\SI{}{\second}}$.}
	\label{fig:rotation_exp}
\end{figure}

 Other than possible radial offsets, the angular alignment of the apertured plane may also be imperfect. Considering the perfect angular alignment case as the scenario in which the apertured plane is perpendicular to the transmitter-receiver axis (i.e., the normal plane), this subsection investigates the error performance when this perpendicularity is broken. Denoting the angular deviation from the normal plane as $\theta$, the cross-sectional view of the topology is as presented in Fig. \ref{fig:rotation_topology}. Using this topology, Fig. \ref{fig:rotation_exp} presents the BER and SINAR vs. $\theta$ curves for fixed $d_a$ and $r_a$ values. Please note that the values for $d_a$ and $r_a$ are fixed in order to isolate the true effect of the angular offsets. Similar to all other evaluated scenarios, the $F_{hit}(t)$ functions are generated using Monte Carlo simulations with $10^5$ molecules, with an incremental time step of $\Delta t = \SI{1e-4}\second$, and a total duration of $\SI{10}\second$. Using these data, the channel coefficients are generated and Monte Carlo BER simulations are performed using bit streams of length $10^6$.

\par The results of Fig. \ref{fig:rotation_exp} show that the error performance of the system worsens with increasing $\theta$, which is also confirmed by the SINAR curve. It is noteworthy that for the system parameters described in Fig. \ref{fig:rotation_exp}, the performance deterioration is observed to be mild, corresponding to one order of magnitude BER increase in approximately $\SI{50}{\degree}$. Overall, although a generalized claim requires more analysis, our initial results suggest that radial offsets are considerably more detrimental to the apertured-plane systems compared to angular imperfections.

\section{Conclusions and Future Work}

\par In this paper, we considered the performance of an MCvD system in the presence of an apertured plane. Either manually placed or naturally occurring, we have found that the overall error performance can actually be improved with an apertured plane, even though such an obstacle decreases received power. We have found that this improvement in performance is mainly due to the ISI mitigation introduced by the apertured plane. We have recognized and reported a trade-off between the ISI mitigation and received signal power, and shown that there exists an optimal aperture size that minimizes BER. In order to quantify the communication efficiency of the overall system, we proposed a novel metric called SINAR and realized it accurately represents the overall BER performance of the system. Acknowledging the trade-off, we have investigated the effects of various topological and communication parameters on the mentioned optimal radius. Lastly, motivated by the practical difficulties in realizing a concentric aperture, we have investigated the effects of misalignments and shown that the BER improvement over the benchmark scenario still holds up to a certain point.

\par Considering the presented work in this paper, future research directions include:
\begin{itemize}
    \item the consideration of absorbing and partially absorbing apertured planes,
    \item investigation of other apertured plane-like obstacles that yield similar ISI mitigation properties,
    \item further analysis of the effects of possible angular rotations of the apertured planes,
    \item scenarios with multiple apertured planes, with concentric and non-concentric alignments,
    \item scenarios where the perfect synchronization assumption does not hold,
    \item and the development of a macro-scale testbed implementation of the considered topological model.
\end{itemize}

\section*{Acknowledgements}

\par This work was partially supported by the Turkish Directorate of Strategy and Budget under the TAM project number 2007K12-873.

\bibliographystyle{IEEEtran}
\bibliography{index_refs_new}

\end{document}